\documentclass[aps, prx, reprint, notitlepage, showpacs, floatfix, superscriptaddress]{revtex4-1}

\usepackage{silence}
\usepackage{amsthm,amsmath,amsfonts,amssymb,verbatim, color}
\usepackage[percent]{overpic}
\usepackage{epsfig,slashed}
\usepackage{epstopdf}
\usepackage{lipsum}
\usepackage{float}
\usepackage{mathtools}
\usepackage[colorlinks=true,citecolor=blue,linkcolor=black,urlcolor=blue]{hyperref}
\usepackage{natbib}
\usepackage[mathscr]{euscript}
\usepackage[export]{adjustbox}

\usepackage[caption=false]{subfig}
\usepackage{commath}
\usepackage{graphicx,bm}

\makeatletter
\newsavebox\myboxA
\newsavebox\myboxB
\newlength\mylenA

\newcommand*\xoverline[2][0.75]{%
    \sbox{\myboxA}{$\m@th#2$}%
    \setbox\myboxB\null% Phantom box
    \ht\myboxB=\ht\myboxA%
    \dp\myboxB=\dp\myboxA%
    \wd\myboxB=#1\wd\myboxA% Scale phantom
    \sbox\myboxB{$\m@th\overline{\copy\myboxB}$}%  Overlined phantom
    \setlength\mylenA{\the\wd\myboxA}%   calc width diff
    \addtolength\mylenA{-\the\wd\myboxB}%
    \ifdim\wd\myboxB<\wd\myboxA%
       \rlap{\hskip 0.5\mylenA\usebox\myboxB}{\usebox\myboxA}%
    \else
        \hskip -0.5\mylenA\rlap{\usebox\myboxA}{\hskip 0.5\mylenA\usebox\myboxB}%
    \fi}
\makeatother

\begin{document}

\newtheorem{theorem}{Theorem}
\newtheorem{property}{Property}
\newcommand{\tr}{\mathop{\mathrm{Tr}}}
\newcommand{\bsigma}{\boldsymbol{\sigma}}
\newcommand{\re}{\mathop{\mathrm{Re}}}
\newcommand{\im}{\mathop{\mathrm{Im}}}
\newcommand{\diag}{\mathrm{diag}}
\newcommand{\sign}{\mathrm{sign}}
\newcommand{\sgn}{\mathop{\mathrm{sgn}}}
\newcommand{\mb}{\bm}
\newcommand{\ua}{\uparrow}
\newcommand{\da}{\downarrow}
\newcommand{\ra}{\rightarrow}
\newcommand{\la}{\leftarrow}
\newcommand{\mc}{\mathcal}
\newcommand{\bs}{\boldsymbol}
\newcommand{\lra}{\leftrightarrow}
\newcommand{\nn}{\nonumber}
\newcommand{\half}{{\textstyle{\frac{1}{2}}}}
\newcommand{\mf}{\mathfrak}
\newcommand{\MF}{\text{MF}}
\newcommand{\IR}{\text{IR}}
\newcommand{\UV}{\text{UV}}

\renewcommand{\i}{\mathop{\mathrm{i}}}
\renewcommand{\b}[1]{{\boldsymbol{#1}}}

\def\II{\hbox{$1\hskip -1.2pt\vrule depth 0pt height 1.6ex width 0.7pt\vrule depth 0pt height 0.3pt width 0.12em$}}

\DeclareGraphicsExtensions{.png}

\title{Protection of qubits by nonlinear resonances}

\author{Rakesh Kumar Saini}
\affiliation{\mbox{UM-DAE Centre for Excellence in Basic Sciences, Vidynagari campus, Mumbai 400098}}

\author{Raman Sehgal}
\affiliation{Nuclear Physics Division, Bhabha Atomic Research Centre, Mumbai 400085, India}

\author{Sudhir R. Jain}
\affiliation{\mbox{Theoretical Nuclear Physics \& Quantum Computing Section} \\
\mbox{Nuclear Physics Division, Bhabha Atomic Research Centre, Mumbai 400085, India}}
\affiliation{\mbox{Homi Bhabha National Institute, Mumbai 400094, India}}
\affiliation{\mbox{UM-DAE Centre for Excellence in Basic Sciences, Vidynagari campus, Mumbai 400098}}

\begin{abstract}
We show that quantized superconducting circuits are non-integrable at the classical level of description, adorned by nonlinear resonances amidst stochastic sea. The spectral fluctuations of these quasi-integrable systems exhibit intermediate behaviour between regularity and chaos. The distribution function of ratios of adjacent spacings, and, nearest-neighbour spacing distribution functions attest to the occurrence of ``mild chaos". Based on these features, we propose criteria for protection of qubits from decoherence which amounts to choosing the parameters of the system in a way that the system resides as close as possible to the elliptic point of the primary nonlinear resonance of the corresponding classical system. 
\end{abstract}
\maketitle

\hypersetup{linkcolor=blue}

Circuit quantum electrodynamics studies the properties of quantum circuits using Josephson junctions coupled to the photon modes of  microwave cavities \cite{apr,haroche}. There is a tremendous interest in the study of these systems for their connections with fundamental problems in open quantum systems, quantum engineering of states, and decoherence \cite{ksb}. One of the important workhorses for realizing a qubit is a transmon - a Josephson tunnel junction shunted by a capacitance. This is a weakly nonlinear oscillator derived from a Cooper pair box \cite{tsai,clarke}. Transmons are playing the same role in a superconducting quantum computer as played by ``harmonic oscillator" and ``particle in a box (billiards)" at the advent of quantum mechanics. The major difference, of course, is that these are nonlinear systems at the classical level. Moreover, their simplicity is misleading as they are manifestations of collective effects of supercondutors \cite{girvin}. Nevertheless, the transitions among the low-lying levels can be given a description in a manner akin to Rabi oscillations in atomic systems \cite{tsai,nakamura}. Thus, we have a nonlinear quantum system with rich physics, termed in common parlance as an ``artificial atom". 

The nonlinear artificial atomic systems are examples of non-integrable systems of a special kind, well-known in the literature as quasi-integrable systems or the KAM (Kolmogorov-Arnold-Moser) systems \cite{kam}.  We characterize these systems by studying the classical phase space as well as  the spectral properties for several configurations in which transmons have been coupled for applications in quantum computing employing superconducting qubits. The fluctuations in their spectra bear properties akin to the systems described by random matrix theory (RMT) for systems with mixed phase space \cite{haake,stockmann,jha,mehta}. To recall, quasi-integrable systems are obtained upon analytic perturbation of an integrable Hamiltonian. As the strength of the perturbation increases, the invariant tori in the phase space break. Eventually, when the last invariant torus is broken, the system becomes chaotic. In its transition to chaos, the phase space is typically comprised of regions termed as stable islands around elliptic points, surrounded by a  stochastic sea. This is the scenario envisaged by the celebrated work of Kolmogorov, Arnold, and Moser (KAM theorem) \cite{kam,ll,arnold}. Some of the spectacular phenomena resulting due to stability provided by nonlinear resonances are  rings of Saturn and  asteroid belts in our Solar system \cite{wisdom,wisdom1}.  Our study presents a connection between  quantum computing \cite{nc,preskill_Q} and quantum chaos \cite{js,srjain1993,gutzwiller}. 

We shall concentrate on inductively coupled transmons \cite{apr} and the much discussed, $0 - \pi$ qubit \cite{bkp}. Pure capacitively coupled transmon system can be shown to be classically integrable; hence, with any perturbation, it loses its integrability. On the other hand, inductively coupled transmons possess a parameter regime where the system gets trapped in a nonlinear resonance. Trapping in the islands and scattering off  the islands has been studied in great detail, classically \cite{neishtadt,cary} and quantum mechanically \cite{lock,backer}. Outside this regime, the resonant region opens up, making the system susceptible to possible undesirable interaction with environment or noise, leading to decoherence. The configuration, termed a $0 - \pi$ qubit was proposed by Kitaev and developed further by Brooks, Kitaev, and Preskill \cite{bkp}; recently, it is experimentally realized \cite{gyenis}. At the classical level, we show that it posseses a compact phase space and the system is trapped in two nonlinear resonances \cite{kam,neishtadt}. On the basis of the behaviour of dynamical  systems possessing trapping regions created due to nonlinear resonances, and the associated quantum chaos \cite{jha}, we propose criteria for protection of qubits. We believe that this proposal will help design new protected configurations for qubits by making sure that there exist deep trapping regions provided by primary nonlinear resonances. While classically forbidden, these trapping regions are accessible quantum mechanically via tunneling. The tunneling rate can be estimated semiclassically on the basis of the change in adiabatic invariant \cite{cary} when a system crosses a separatrix. 

Let us propose criteria which would help protection of qubits from decoherence or noise. 
\begin{enumerate}
    \item The system may only be weakly nonlinear, belonging to the class of quasi-integrable systems at the classical level of description, or, what are known as KAM  systems. The perturbations to the classically integrable component may be smooth.
    \item Due to this first point, there would exist a heierarchy of islands of stability in the classical phase space, the primary resonance being the largest. The system parameters may be chosen so that the system sits close to the elliptic point of the primary island. This would provide a natural barrier to any external disturbance, which has to tunnel to reach the system. 
    \item At the level of quantum description, the nearest-neighbour level spacing distribution is asked to remain close to Poisson distribution, with little (if any) contribution from various distributions with level repulsion in random matrix theory.
    \item If the available classical phase space is compact, then there will be even stronger protection.   
\end{enumerate}

Let us recall that in the transmon regime, the charging energy is much smaller than the Josephson energy, a fixed frequency transmon being governed by a Hamiltonian \cite{apr}: $H = 4E_C N^2 - E_J \cos \phi $. The number operator, $N$ and superconducting phase difference, $\phi$ are conjugate variables, satisfying $[\phi , N] = i$. To generate entanglement between individual quantum systems, it is necessary to engineer an interaction. We consider inductive coupling \cite{apr} between two transmons (Fig. 1). The Hamiltonian of the system is
\begin{alignat}{1}\label{eq:inductive}
    H &= \sum_{i=1,2} \left[ 4E_{C_{i}}N_i^2 - E_{J_{i}}\cos \phi _i\right] \nonumber \\ &+ M_{12}I_{C_{1}}\sin \phi_1 I_{C_{2}}\sin \phi_2.
\end{alignat}
To explain our criteria for protection, we shall delve into necessary details at the classical level \cite{ll}; we begin by stating the resonance condition,
\begin{alignat}{1}
m\omega_{1}-n\omega_{2} &= m \frac{d\phi_1}{dt}-n\frac{d\phi_2}{dt} \nonumber \\ &= m\frac{\partial H}{\partial N_1} - n\frac{\partial H}{\partial N_2} = 0, \nonumber \\ 
mE_{C_{1}} N_1 &= n E_{C_{2}} N_2.
\end{alignat}
where $m,n$ are integers. We perform a canonical transformation to examine the behaviour of the system close to the region in phase space where the resonance condition is satisfied. The mapping, ($N_1,N_2,\phi_1,\phi_2$) $\mapsto$ ($R,J,\phi,\psi$) is effected by employing the generating function $W = (m \phi_1 - n \phi_2)R - (l_1 \phi_1 -l_2 \phi_2 )J$, where $l_1,l_2$ are integers such that $m l_2 -n l_1 =1$. Using the transforming equations, we obtain
\begin{alignat}{1}
    m E_{C_{1}} (m R -l_1 J) &=n E_{C_{2}}(-n R + l_2 J) \nonumber \\
    R = R_{res}(J) &= \frac{(m \ l_1 \ E_{C_{1}} + n \ l_2 \ E_{C_{2}})}{(m^2 E_{C_{1}} + n^2 E_{C_{2}}) } J.
\end{alignat}
The new Hamiltonian is
\begin{alignat}{1}
    H(R,J,\phi,\psi) &= 4 E_{C_{1}} (m R - l_1 J)^2 +  4 E_{C_{2}} (-n R + l_2 J)^2  \nonumber \\ & - E_{J_{1}} \cos(l_2 \phi + n \psi)  - E_{J_{2}} \cos(l_1 \phi + m \psi) \nonumber \\ & + \beta_{12} \sin(l_1 \phi + m \psi) \sin(l_2 \phi+ n \psi)
\end{alignat}
where $\beta_{12} = M_{12} I_{C_{1}} I_{C_{2}}$. As $\frac{d \phi}{dt} = 0$ at resonance,  $\phi$ is a slow variable, and $\psi$ is a faster variable (near resonance). On integrating over the fast variable. and noting that the conjugate variable $J$ is an integral of motion, the Hamiltonian reads as
\begin{alignat}{1}
H_1(R,J,\phi) &= \frac{1}{2 \pi}\int_{-\pi}^{\pi} H(R,J,\phi,\psi) d \psi \nonumber \\ & = E_{C_{1}}(4 J^2 l_{1}^{2} - 8 J l_1 n R + 4 n^2 R^2) \nonumber \\ & + E_{C_{2}}(4 J^2 l_{2}^{2} - 8 J l_2 n R + 4 n^2 R^2) \nonumber \\ & +  (\beta_{12}/2) \cos((l_1-l_2)\phi)
\end{alignat}
for $m=n$; the general expression can be  found but is quite cumbersome to be given here. We apply yet another canonical transformation ($R,\phi$) $\mapsto$ ($P,\phi$) employing the generating function, $W' = (P + R_{res}(J) )\phi$ with new variable $P = R -R_{res}$. We obtain the resonant Hamiltonian:
\begin{alignat}{1}
    H_{\rm res}(P,\phi) &= 4(m^2 E_{C_{1}}+n^2 E_{C_{2}}) P^{2} \nonumber \\ &+ \frac{\beta_{12}m}{2} \cos[(l_2 -l_1) \phi] + \Lambda(J)
\end{alignat}
where 
\begin{equation}
\Lambda(J)= \frac{4 E_{C_{1}}E_{C_{2}}(l_{2}^{2} m^2- 2 l_1 l_2 m n + l_{1}^{2} n^2)}{m^2 E_{C_{1}}+ n^2 E_{C_{2}}} J^2.    
\end{equation}
The values of the parameter that we have taken are: $E_{C_{1}}=0.002$ GHz, $E_{C_{2}}=0.003$ GHz, $E_{J_{1}}=E_{J_{2}}= 1 $GHz and $m:n = 1:1$ as primary resonance. Now we are ready to exhibit the phase space surface corresponding to resonance condition $\frac{N_1}{N_2}= \frac{E_{C2}}{E_{C1}}= 3/2$.

\begin{figure*}[hbt!]
\subfloat[\label{classicalmcop001}]{%
  \includegraphics[width=0.3\textwidth]{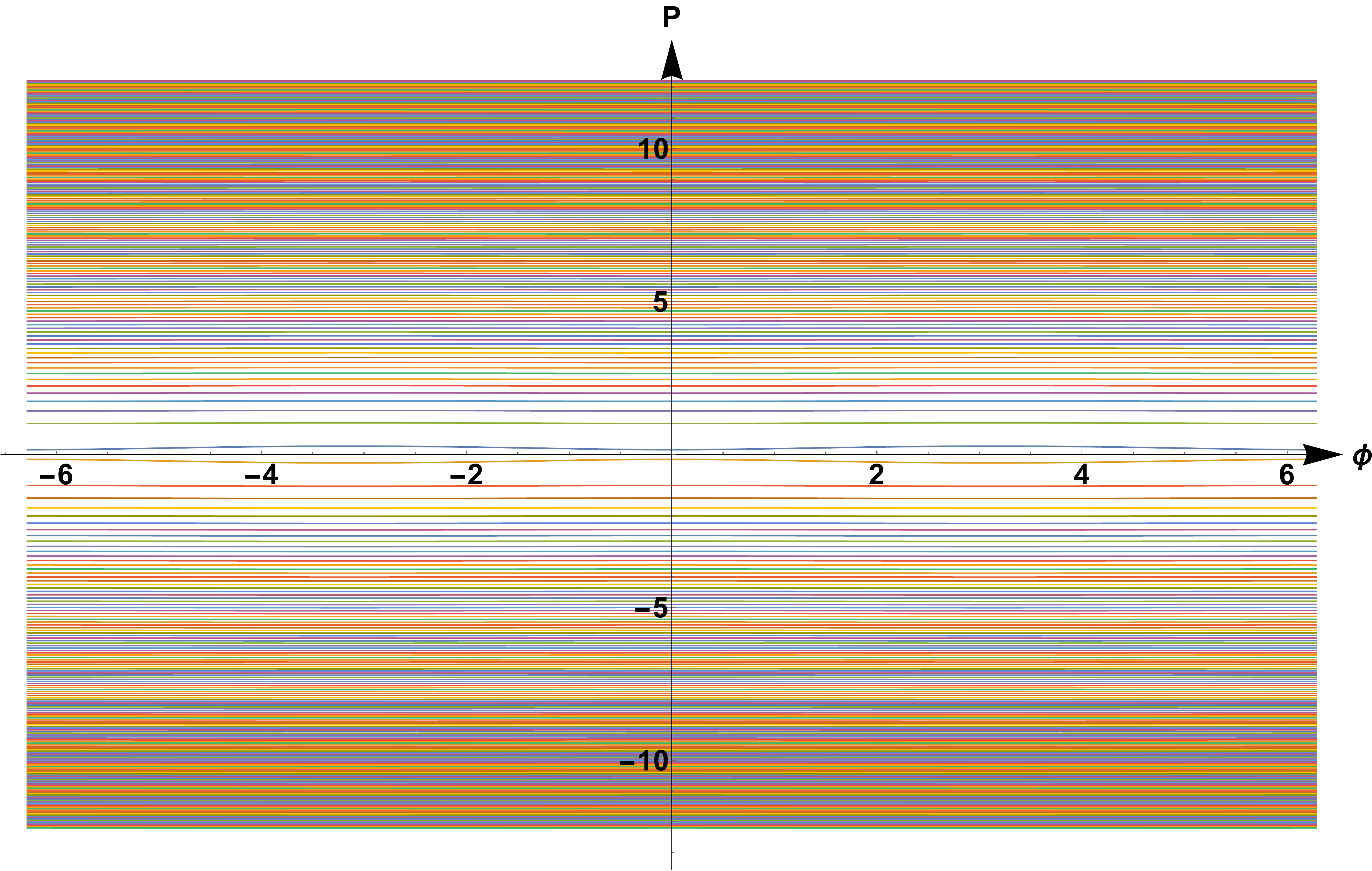}  %
}\hfill
\subfloat[\label{indcoup001}]{%
\includegraphics[width=0.3\textwidth]{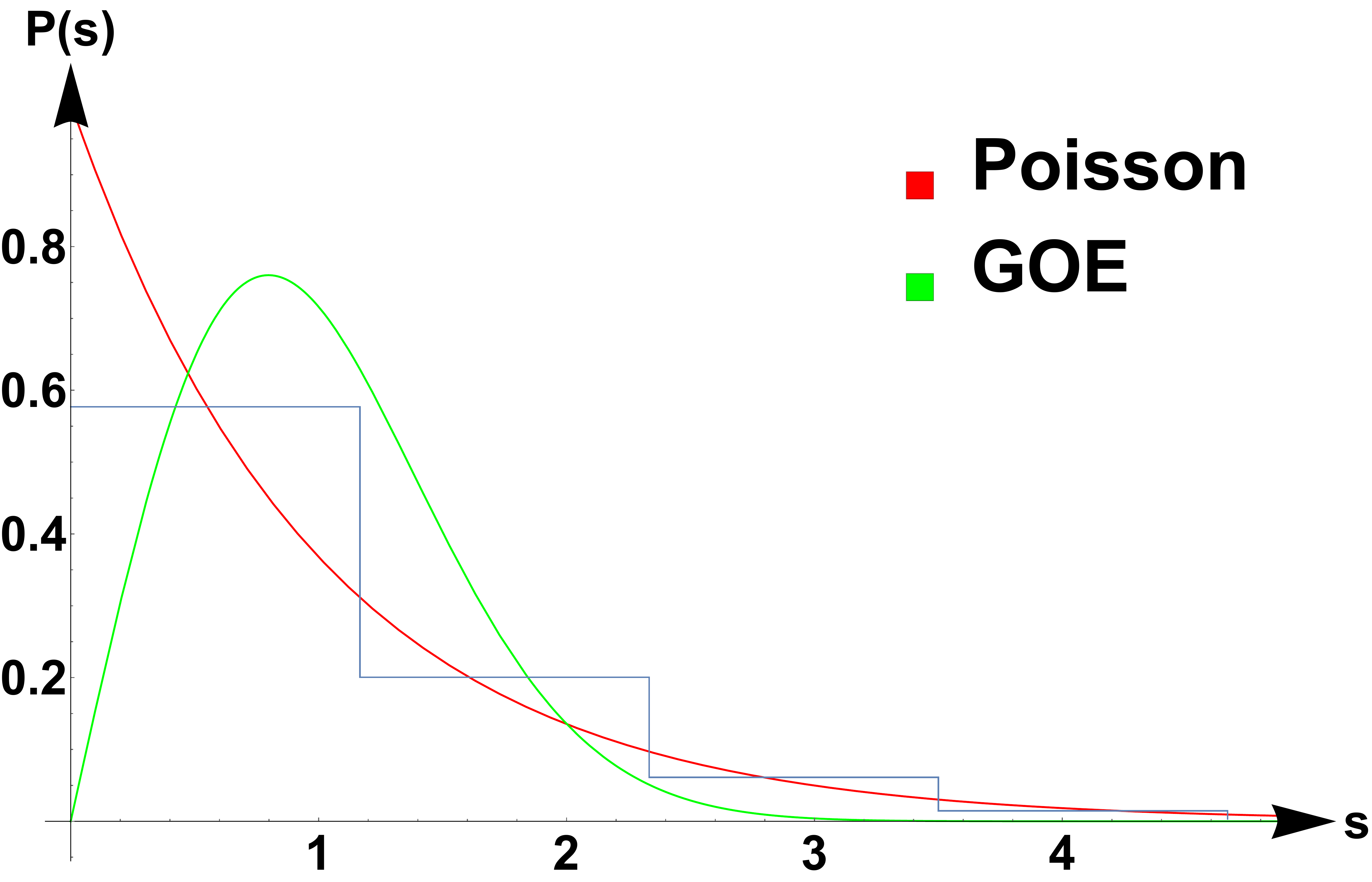} %
}\hfill
 \subfloat[\label{brodyindcoup1}]{%
\includegraphics[width=0.3\textwidth]{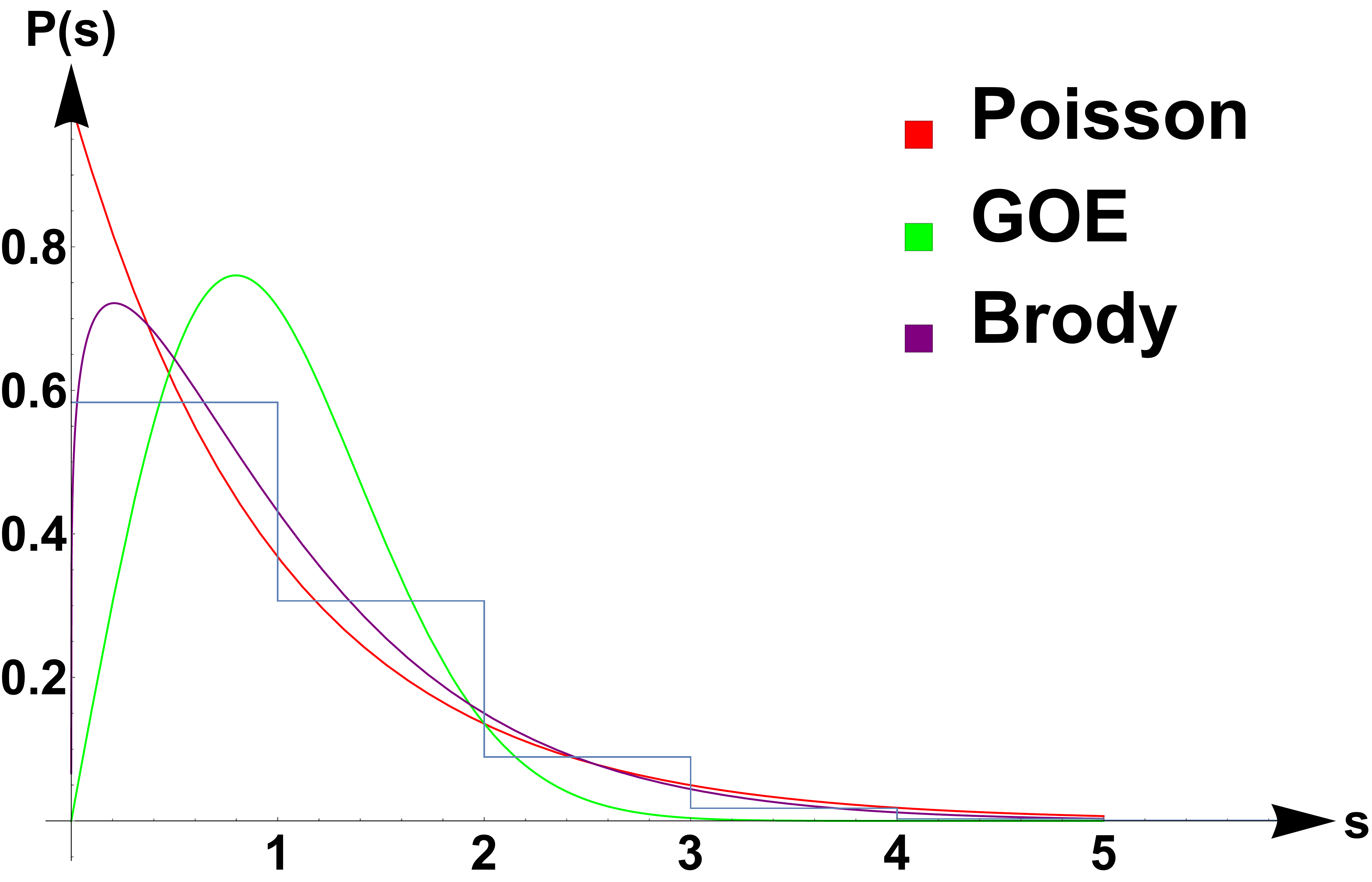}%
}
 \\
 \subfloat[\label{classicalmcop1}]{%
 \includegraphics[width=0.4\textwidth]{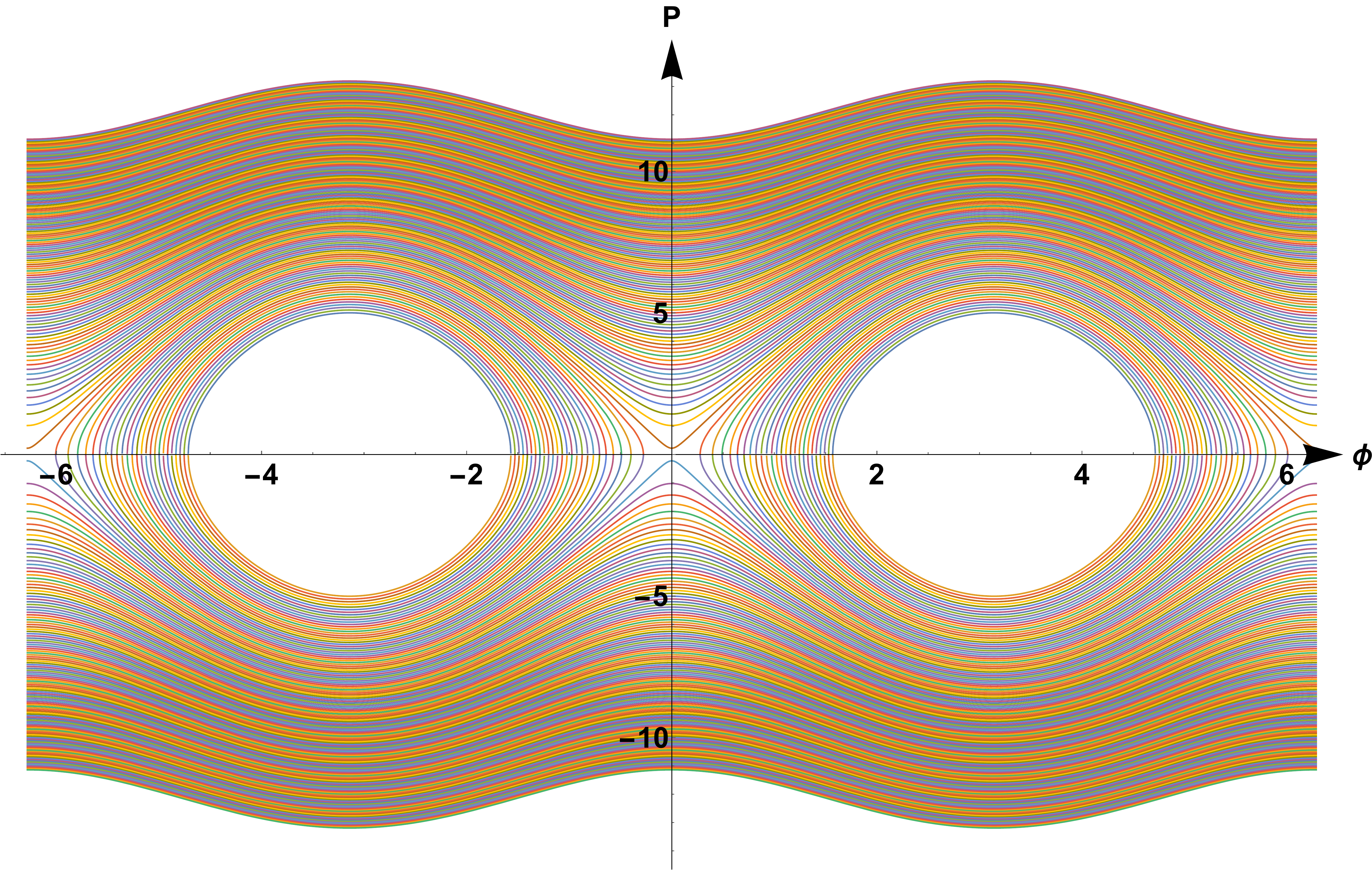} %
}\hfill
\subfloat[\label{tunnelingindcoup}]{%
\includegraphics[width=0.4\textwidth]{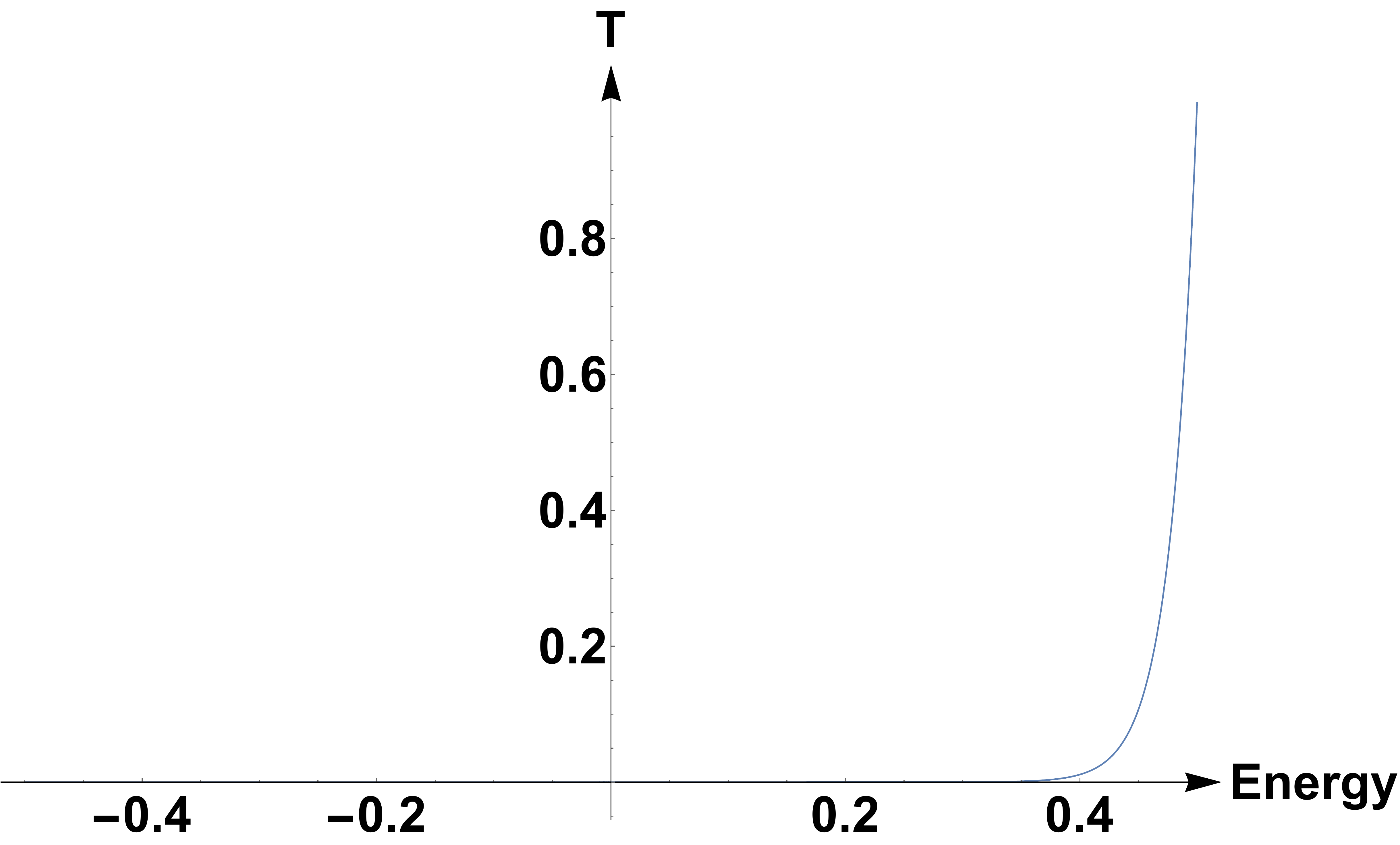}
}
\\
\subfloat[\label{groundlevelindcoup}]{%
  \includegraphics[width=0.4\textwidth]{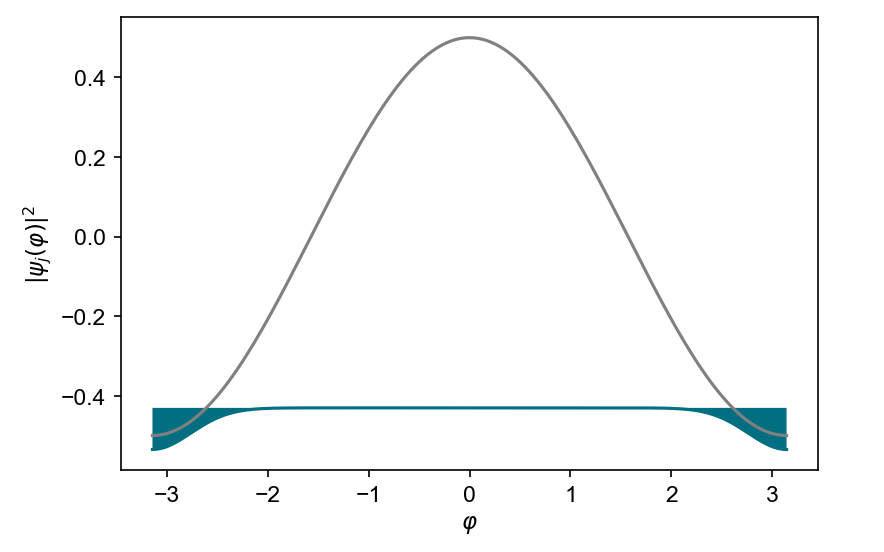}%
}
\hfill
\subfloat[\label{firstexclevelindcoup}]{%
\includegraphics[width=0.4\textwidth]{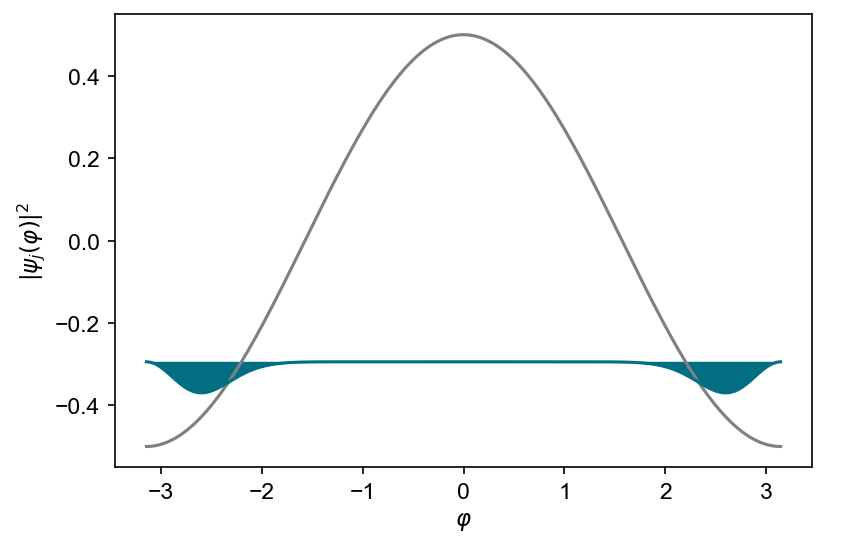}
}
\caption{In {\ref{classicalmcop001}) \& \ref{classicalmcop1})}, we have the phase space plot of Hamiltonian near resonance (1:1), respectively for $\beta=0.001$ and $\beta=1$. For $\beta=1$ in \ref{groundlevelindcoup}) \& \ref{firstexclevelindcoup}), the square of the wavefunction is plotted respectively for the ground state and the excited state of the resonant Hamiltonian, with the cosine potential for reference. Fig. \ref{tunnelingindcoup}) is the estimation of probability of tunneling of a wavepacket through the separatrix, plotted against the energy of the system. In 
Figs \ref{indcoup001}) \& \ref{brodyindcoup1}), we present the probability distribution functions of the nearest-neighbour spacing between the energy levels of the Hamiltonian, (1) for the values of parameter, $\beta_{12}=0.001$ and $\beta_{12}=1$ to show the quantum signature of weak nonlinearity and the associated non-integrability as $\beta_{12}$ increases. The red line represents the Poisson distribution, the green line represents the Wigner Distribution, and the purple line represents how well the Brody distribution fits with the Brody parameter=$0.174$ (this is calculated using the least square fit of the curve).}
\label{}
\end{figure*}

In addition to the classical dynamics and a semiclassical estimate (Fig. 1), we now turn to a discussion concerning the nature of the energy spectrum of Hamiltonian given by (1). A small departure from integrability is captured by the spectral fluctuations, owing to the well-known studies in quantum chaos. These are based on the celebrated trace formulae \cite{gutzwiller,jain,brack}. There has been a lot of work on establishing connections between classical non-integrability and the fluctuations in the energy spectra about a mean level density \cite{bgs,berry,date1995,jain1997,muller}, and, a large number of applications to many-body physics \cite{carlo,kaur1,kaur2}. Broadly, level clustering (repulsion) is observed in quantum systems with classically integrable (nonintegrable) dynamics. For systems with mixed phase space, the KAM systems, the fluctuations exhibit a mixture of trends. There are many statistical measures for quantifying the fluctuations, we consider here the most popularly studied, nearest-neighbour level-spacing distribution function (Figs 1b), 1c)). For integrable systems, this is a Poisson distribution \cite{berry_tabor} whereas for chaotic systems, the distribution is of the Wigner form corresponding to the Gaussian Orthogonal Ensemble of random matrices \cite{haake,stockmann}. For KAM systems, the distribution is intermediate, given by the Berry-Robnik form \cite{br}. For this case, the degree of level repulsion increases as the fraction of chaotic region increases. We would like the system to be far from chaos, but nevertheless take advantage of ``immunity" provided by the primary island of stability in phase space. The system can be designed to sit deep into the island, and it provides a barrier for the system to escape, as explained above.  As seen in Fig. 1g), for energies close to the elliptic point of the primary island, the tunneling is negligible. This corresponds to the choice of experimental parameters respecting the resonance condition.  

For \eqref{eq:inductive} with values of the coupling strength, $\beta_{12}$,  from 0.001 to 1, the ratios of adjacent spacings are also statistically analyzed. The classical Hamiltonian corresponding to the \eqref{eq:inductive} supports nonlinear resonances (Figs. 1a), 1d)). Thus, there are regions which have islands of stability around elliptic points, surrounded by a stochastic sea. Thus, the classically allowed phase space volume has a fraction corresponding to regular dynamics and the rest supporting chaotic dynamics. The Brody distribution \cite{brody} was shown  to describe the fluctuations:
\begin{alignat}{1}
P_{\rm B}(s) &= \nu (q+1) s^q \exp(-\nu s^{q+1})
\end{alignat}
where, $\nu = \left[\Gamma(\frac{q+2}{q+1})\right]^{q+1}$ and $q$ is the so-called Brody parameter. For  $q=0$, The Brody distribution interpolates between the Poisson distribution ($q=0$) and Wigner distribution ($q=1$).

\begin{figure*}[hbt!]
\subfloat[\label{indm=.001,k=1,r}]{%
\includegraphics[width=0.2\textwidth]{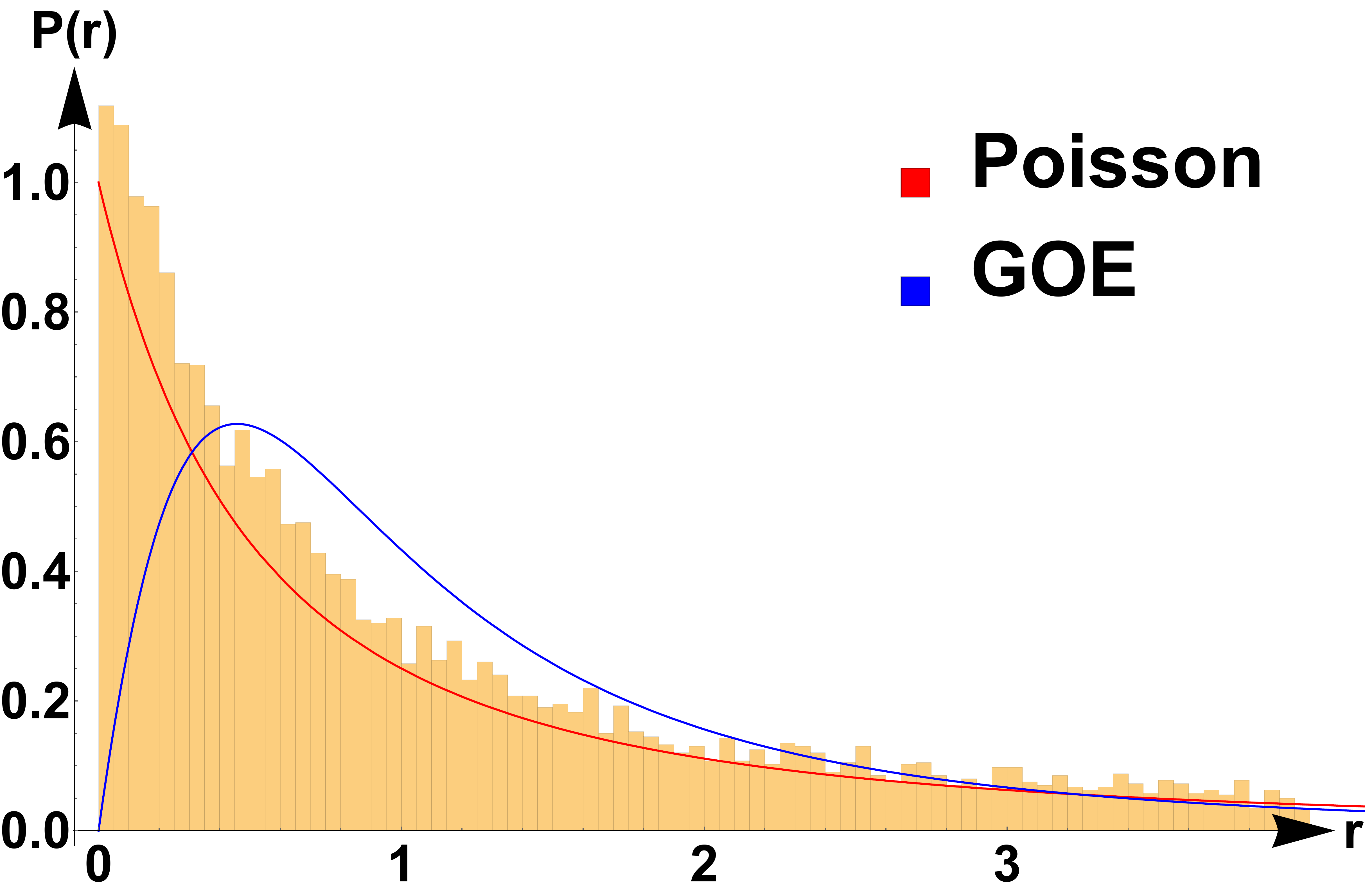}  %
}\hfill
\subfloat[\label{m=1spacingintermediate}]{%
  \includegraphics[width=0.2\textwidth]{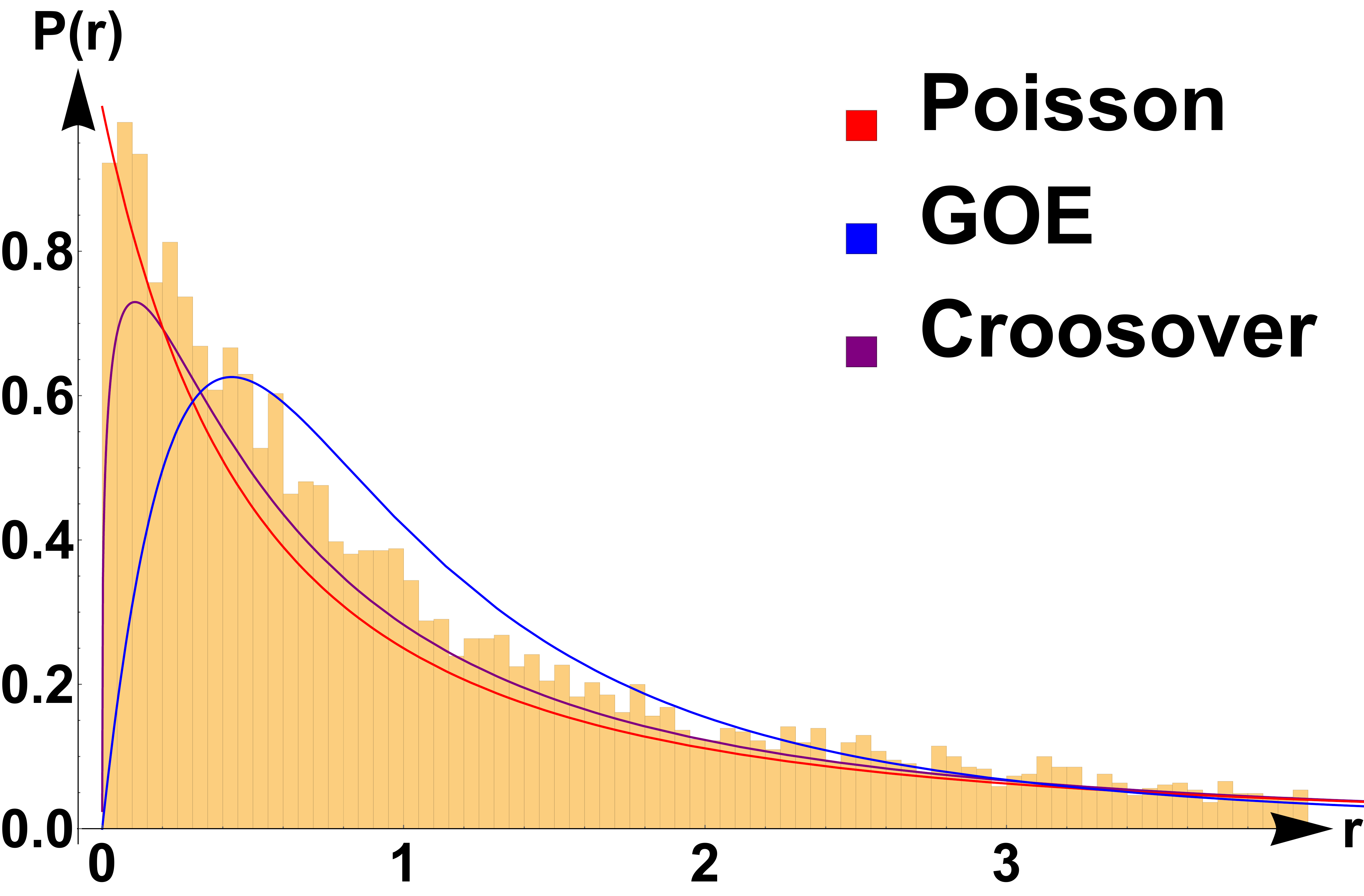}%
}\hfill
\subfloat[\label{indm=.001,k=2,r}]{%
\includegraphics[width=0.2\textwidth]{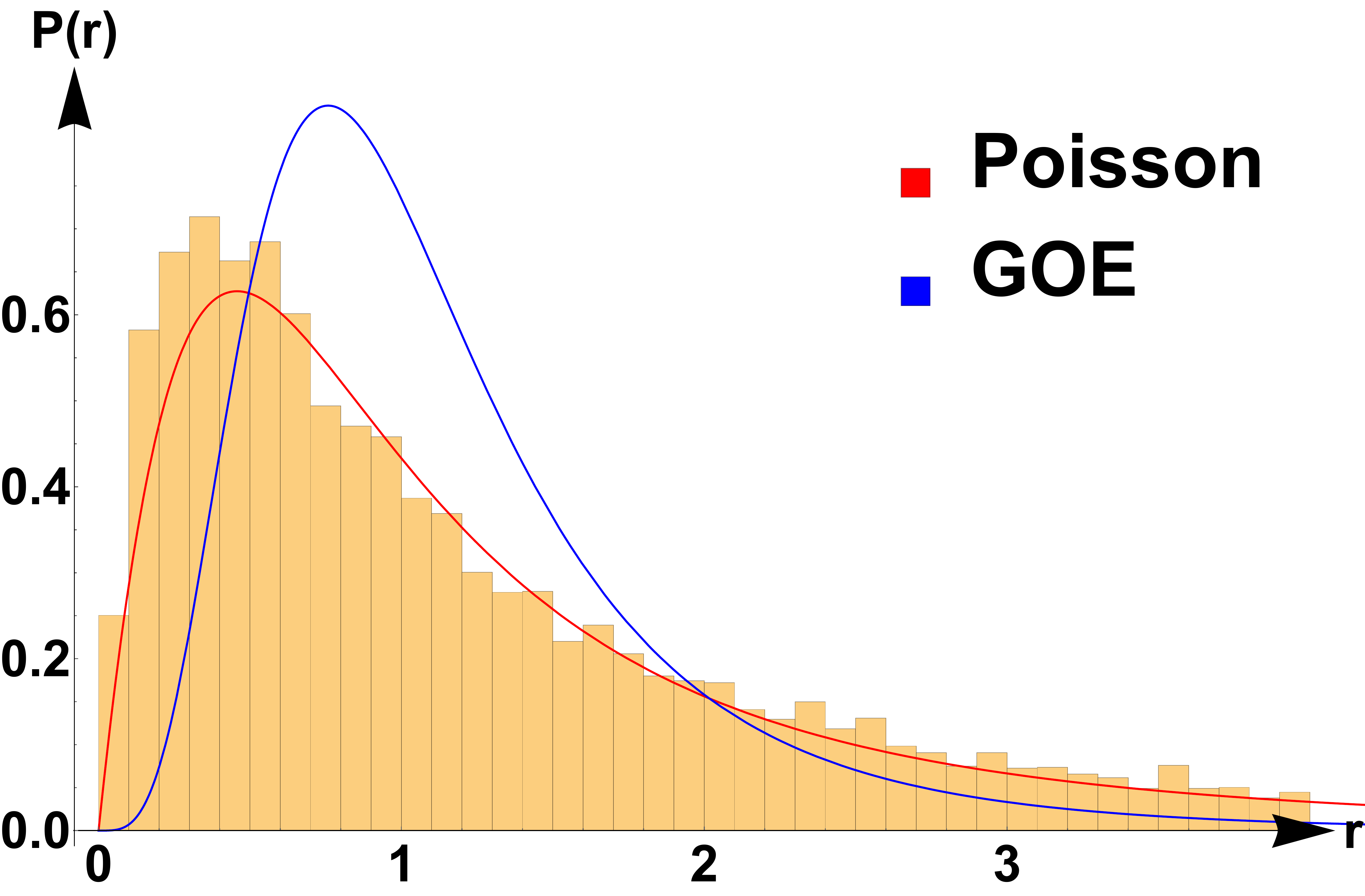}
}\hfill
\subfloat[\label{indm=1,k=2,r}]{%
    \includegraphics[width=0.2\textwidth]{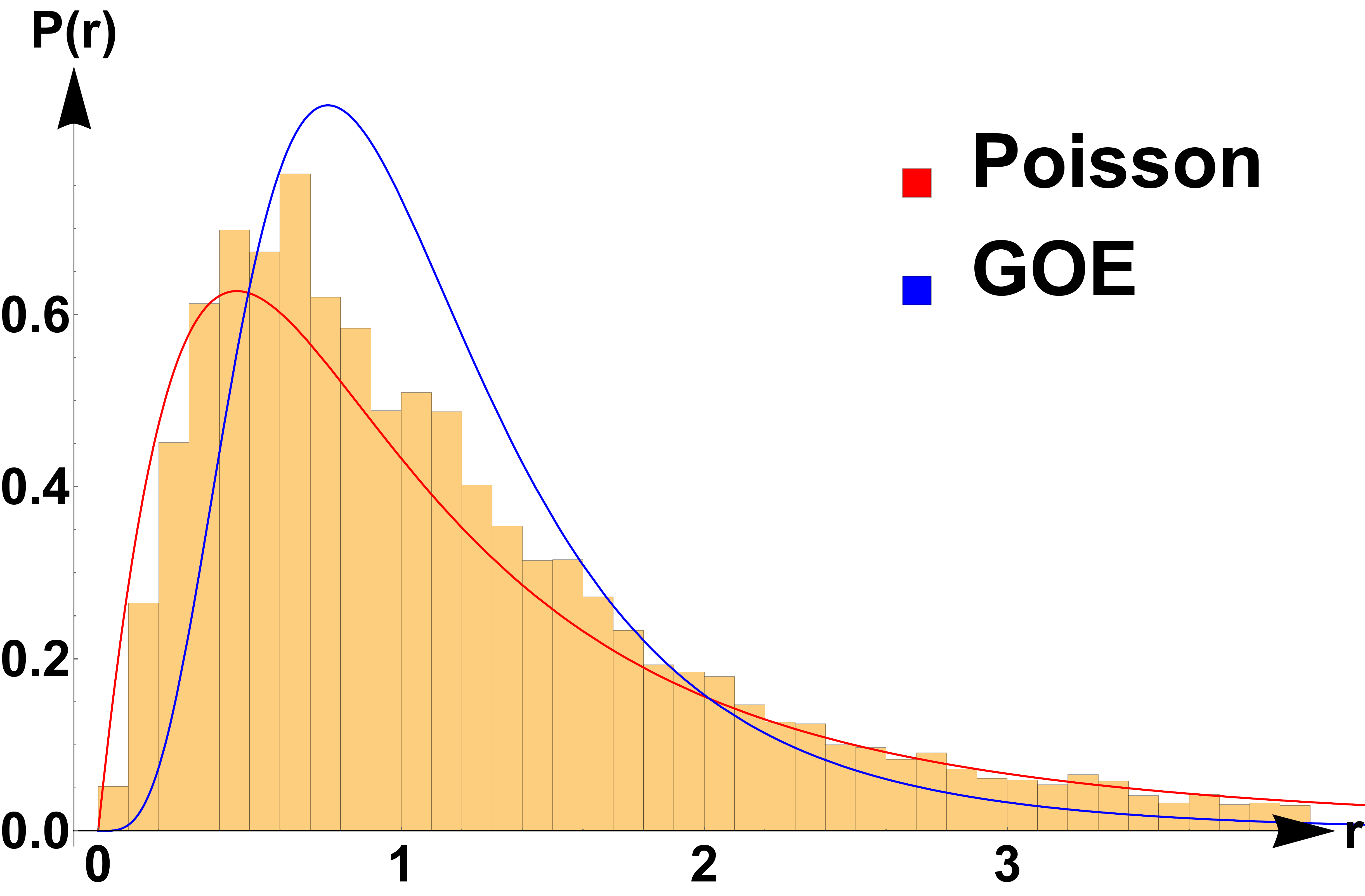}%
}
\caption{In {\ref{indm=.001,k=1,r} \& \ref{m=1spacingintermediate}}, the histogram of ratios of adjacent spacing whereas in {\ref{indm=.001,k=2,r} \& \ref{indm=1,k=2,r}} the histogram of ratios of next nearest neighbour spacing is  plotted respectively for the $\beta_{12}=0.001$ and $\beta_{12}=1$. The red line represents the Poisson distribution ($P(r) = \frac{1}{(1+r)^2}$ distribution for ratios of nearest neighbour spacing and $P(r) = \frac{27}{8}\frac{r + r^2}{(1+r+r^2)^{2.5}}$ for ratios of next nearest neighbour spacing) and blue line represents the GOE or Wigner Distribution ($P(r) = \frac{27}{8}\frac{r + r^2}{(1+r+r^2)^{2.5}}$ distribution for ratios of nearest neighbour spacing and $P(r) = 100.5 \frac{(r + r^2)^4}{(1+r+r^2)^{7}}$ for ratios of next nearest neighbour spacing) and purple line in \ref{m=1spacingintermediate} represents the crossover distribution \cite{corps}
fitted at the parameter value, 0.17.}
\end{figure*} 

\begin{figure*}[hbt!]
\subfloat[\label{zeropi}]{%
  \includegraphics[width=0.3\textwidth]{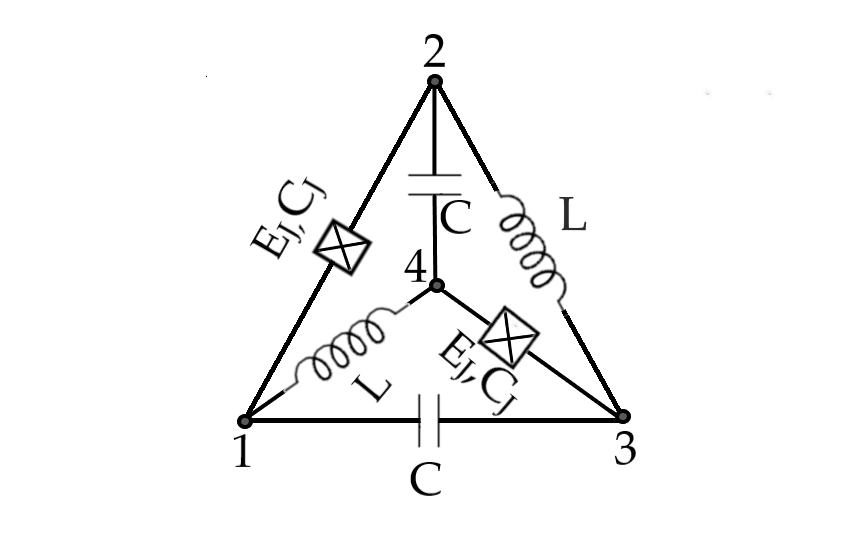}  %
}\hfill
\subfloat[\label{thetamode}]{%
  \includegraphics[width=0.3\textwidth]{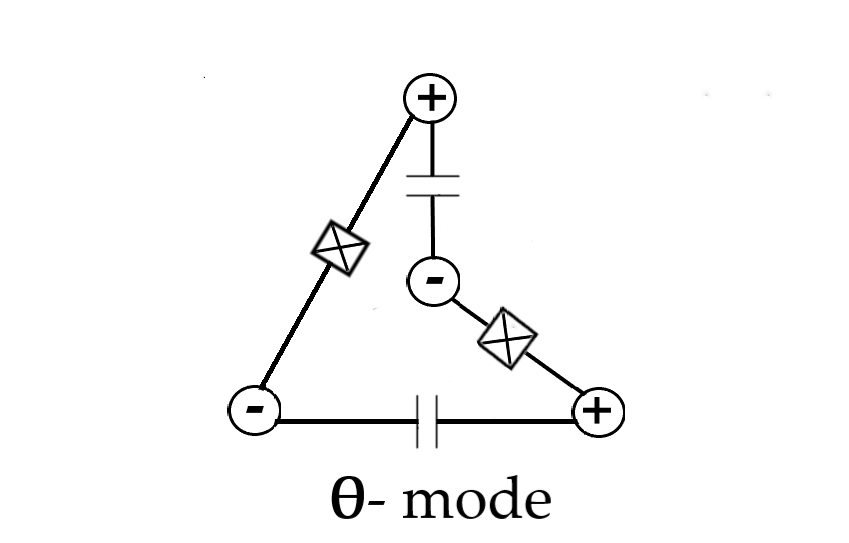}%
}\hfill
\subfloat[\label{phimode}]{%
 \includegraphics[width=0.3\textwidth]{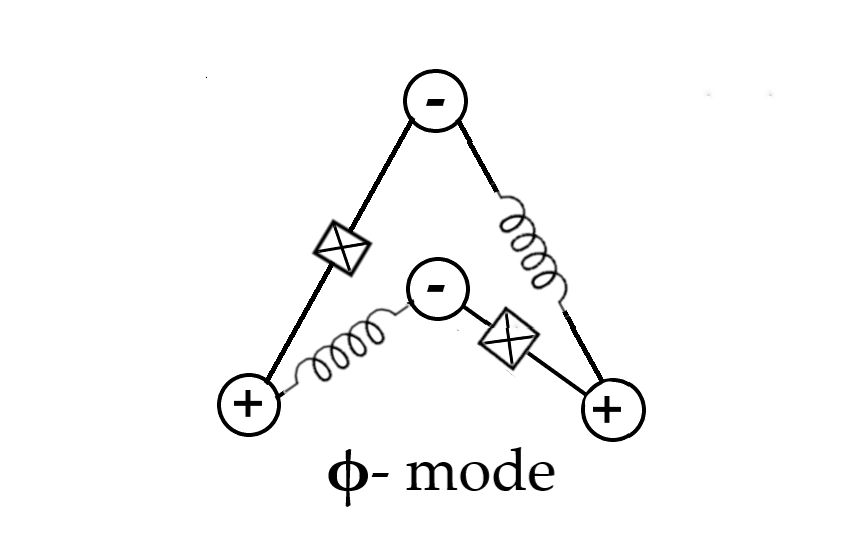} %
}
\caption{\ref{zeropi}) The circuit diagram of the $0-\pi$ qubit. There are four nodes, large capacitors and superinductors, \ref{thetamode}) and \ref{phimode}) Normal modes considered here with signs at the nodes representing their amplitudes.}
\end{figure*}

The ($0-\pi $) qubit consists of identical pairs of small Josephson junctions, large shunting capacitors and superinductors, organized in a small closed loop geometry with four nodes (see Fig. \ref{zeropi}).  It can be shown that this circuit has four degrees of freedom, denoted by (say) $\theta, \phi, \zeta, \Sigma$ modes \cite{gyenis}, corresponding to linear combinations of phase difference between the superconducting order parameter across the elements in the circuit. The $\phi$ and $\theta$ modes describe qubit degrees of freedom of the circuit with the two-mode Hamiltonian \cite{gyenis},
\begin{alignat}{1}\label{eq:zeropi}
H_{0-\pi} &= 4E_C^{\theta} (n_\theta - n_g^{\theta})^2 + 4E_C^{\phi} n_\phi ^2 \nonumber \\ 
 &- 2E_J \cos \theta \cos (\phi - \pi \Phi_{ext}/\Phi_0) + E_L \phi^2
\end{alignat}
where $\Phi_0 = h/2e$, $\Phi_{ext}$ is the external magnetic flux and $n_g^{\theta}$ is the offset-charge bias due to electrostatic environment.

To unravel the phase space structure of the classical Hamiltonian \eqref{eq:zeropi}, we follow the same procedure as above and state the resonance condition between $\theta$- and $\phi$- modes:
\begin{alignat}{1}
m \omega_1 - n \omega_2 &= m \frac{d \theta}{dt} - n \frac{d \phi}{dt} \nonumber \\ & = m \frac{\partial H}{\partial n_{\theta}} - n \frac{\partial H}{\partial n_{\phi}} \nonumber \\
&= m E_C^{\theta}(n_{\theta} - n_g) - nE_C^{\phi}n_{\phi}= 0.
\end{alignat}
Canonically transforming the Hamiltonian ($n_{\theta},n_{\phi},\theta,\phi$) $\mapsto$ ($R,J,\Phi,\psi$) using the generating function $W = (m \theta - n \phi)R - (l_1 \theta -l_2 \phi )J$, where $l_1,l_2$ are integers such that $m l_2 -n l_1 =1$.
\begin{eqnarray}
n_{\theta} &= \frac{\partial W}{\partial \theta} = m R - l_1 J,  \quad
n_{\phi} = \frac{\partial W}{\partial \phi} = -n R + l_2 J, \nonumber\\
\Phi &= \frac{\partial W}{\partial R} = m \theta - n \phi , \quad
\psi = \frac{\partial W}{\partial J} = -l_1 \theta + l_2 \phi. 
\end{eqnarray}
The resonance condition in the new coordinates is
\begin{alignat}{1}
&m E_C^{\theta} (mR - l_1 J -n_g) - n E_C^{\phi} (-nR +l_2 J) = 0, \nonumber \\
&R_{res} = \frac{( m l_1 E_C^{\theta} +n l_2 E_C^{\phi})J + m n_g}{m^2 E_C^{\theta}+ n^2 E_C^{\phi}}
\end{alignat}
being the resonant Hamiltonian. Near resonance, integrating over the fast variable ($\psi$):
\begin{alignat}{1}
H_1(R,J,\phi) &= \frac{1}{2 \pi}\int_{-\pi}^{\pi} H(R,J,\phi,\psi) d \psi \nonumber  \\ & = 4 E_C^{\theta} (J+n_g-R)^2 + 4 E_C^{\phi} (R-2 J)^2
\nonumber \\ & -E_J \cos (\Phi +\Phi_{ext})+E_L \Phi^2+\frac{\pi^2 E_L}{3}.
\end{alignat}
The expression for general $m,n$ is found to be too cumbersome to be presented; moreover, for our purpose, the primary resonance $(m:n=1:1)$, is relevant. After another canonical transformation, ($R,\Phi$) $\mapsto$ ($P,\Phi$) using the generating function $W' = (P + R_{res}(J) )\Phi$ where new variable $P = R - R_{res}$,  We get the Hamiltonian near resonance:
\begin{alignat}{1}
H(P,\Phi) &= 4(E_C^{\theta}+E_C^{\phi}) P^{2} \nonumber \\ &+ \frac{(E_L(\pi^{2}+3 \Phi^{2})- 3 E_J \cos({\Phi_{ext}+ \Phi }))}{3} + \Lambda(J) 
\end{alignat}
where $\Lambda(J) =\frac{4 E_C^{\theta} E_C^{\phi}(n_g - J)^{2}}{E_C^{\theta}+E_C^{\phi}} $. We have taken the parameters  \cite{gyenis} where $E_{C}^{\theta}$= 92 MHz, $E_{C}^{\phi}$ = 1.14 GHz, $E_{J}$ =  6 GHz, $E_{L}$ = 0.38 GHz, $n_{g}$ = 0. The resonance condition for $1:1$ resonance is $\frac{n_\theta}{n_\phi}= \frac{E_{C}^{\phi}}{E_{C}^{\theta}}$ $\simeq $ 12.39. In Fig. 4a) and c), we observe that the classical phase space surrounding the resonance condition is compact. The corresponding potentials and the states thereof are shown in Figs 4b) and d). Figs 4e) - 4j) show the spectral fluctuations of the Hamiltonian operator, $H_{0-\pi}$. These results satisfy our criterion 3. where we would like the system to be only weakly non-integrable.  

\begin{figure*}[hbt!]
\subfloat[\label{0-piclassical0}]{%
  \includegraphics[width=0.4\textwidth]{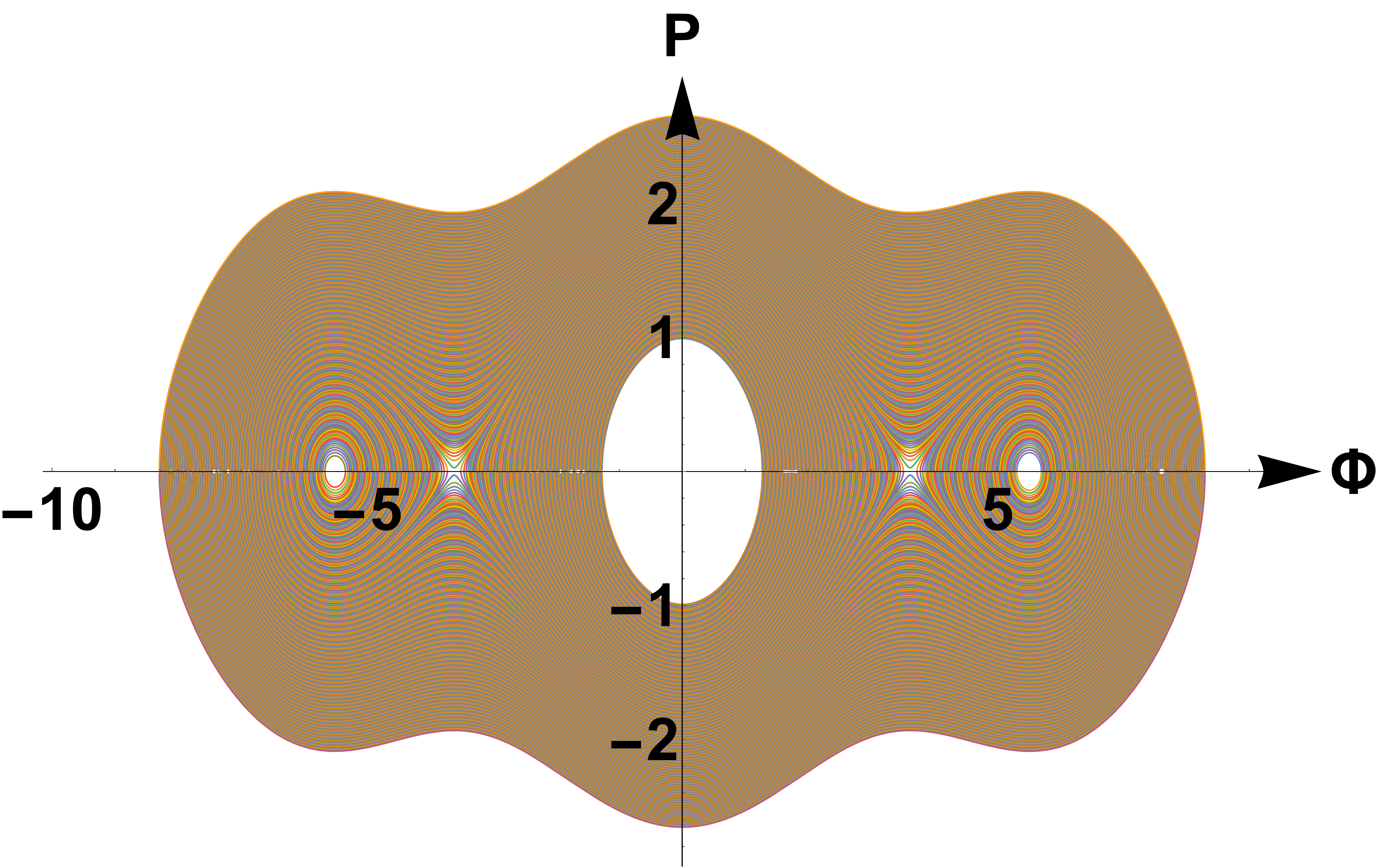}  %
}\hfill
\subfloat[\label{1to10energylevel0-piqubitext0}]{%
  \includegraphics[width=0.4\textwidth]{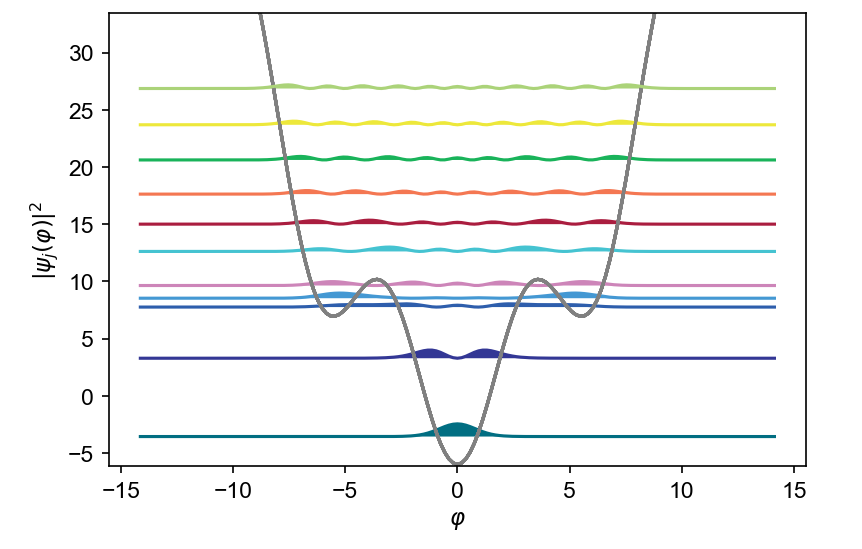}%
} \\
\subfloat[\label{extpiclassical0-pi}]{%
 \includegraphics[width=0.4\textwidth]{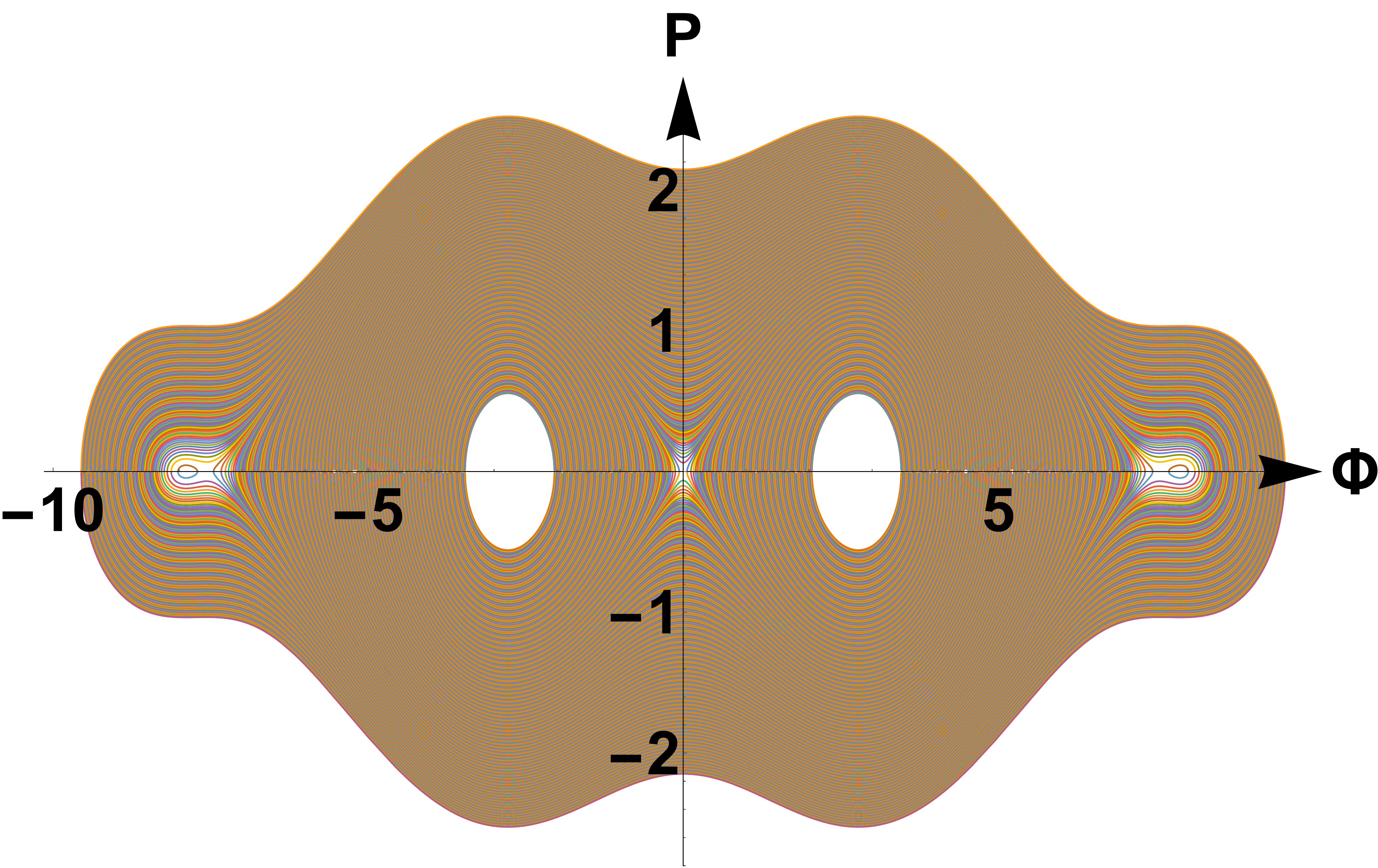} %
}\hfill
\subfloat[ \label{1to10energylevel0-piqubitextpi}]{%
\includegraphics[width=0.4\textwidth]{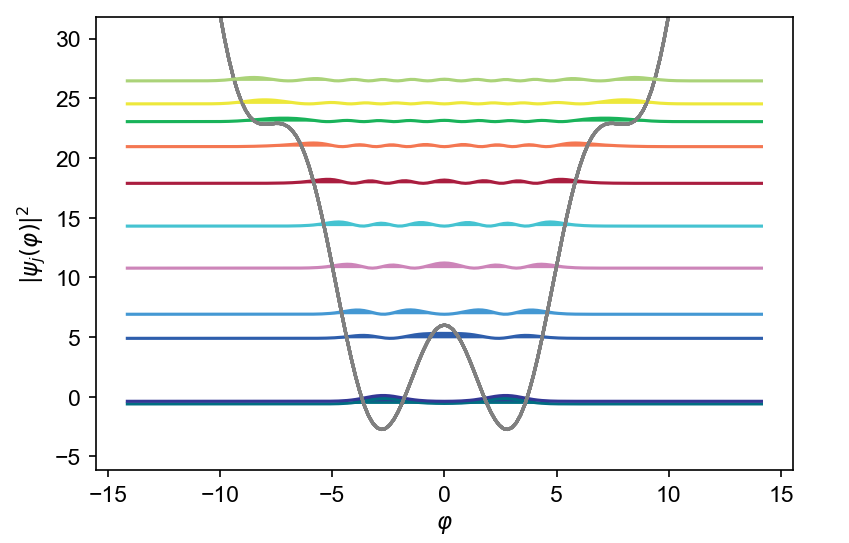}%
}
 \\
 \subfloat[\label{0piext0}]{%
  \includegraphics[width=0.3\textwidth]{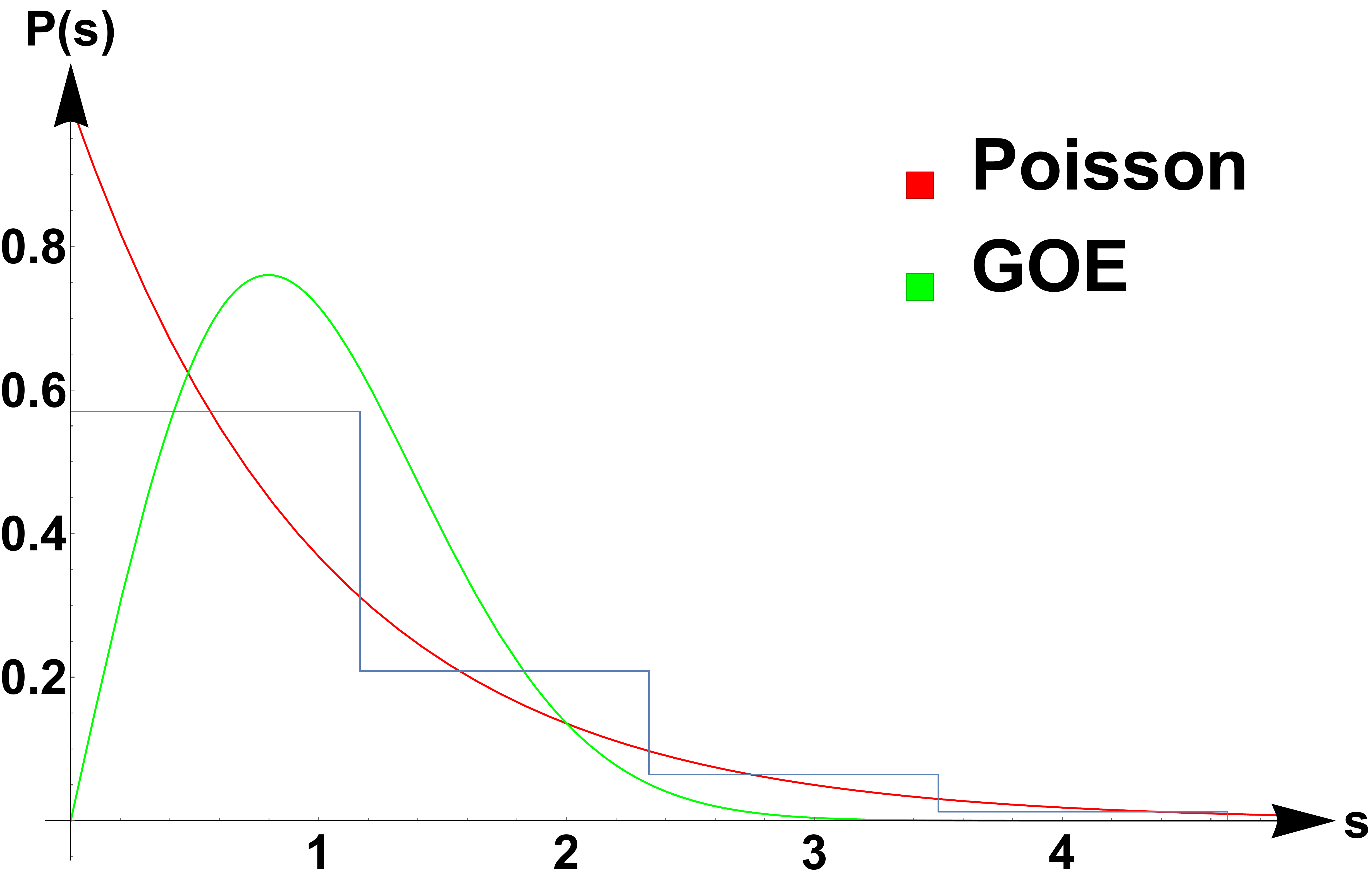}%
}\hfill
\subfloat[\label{0piext0,k=1,r}]{%
\includegraphics[width=0.3\textwidth]{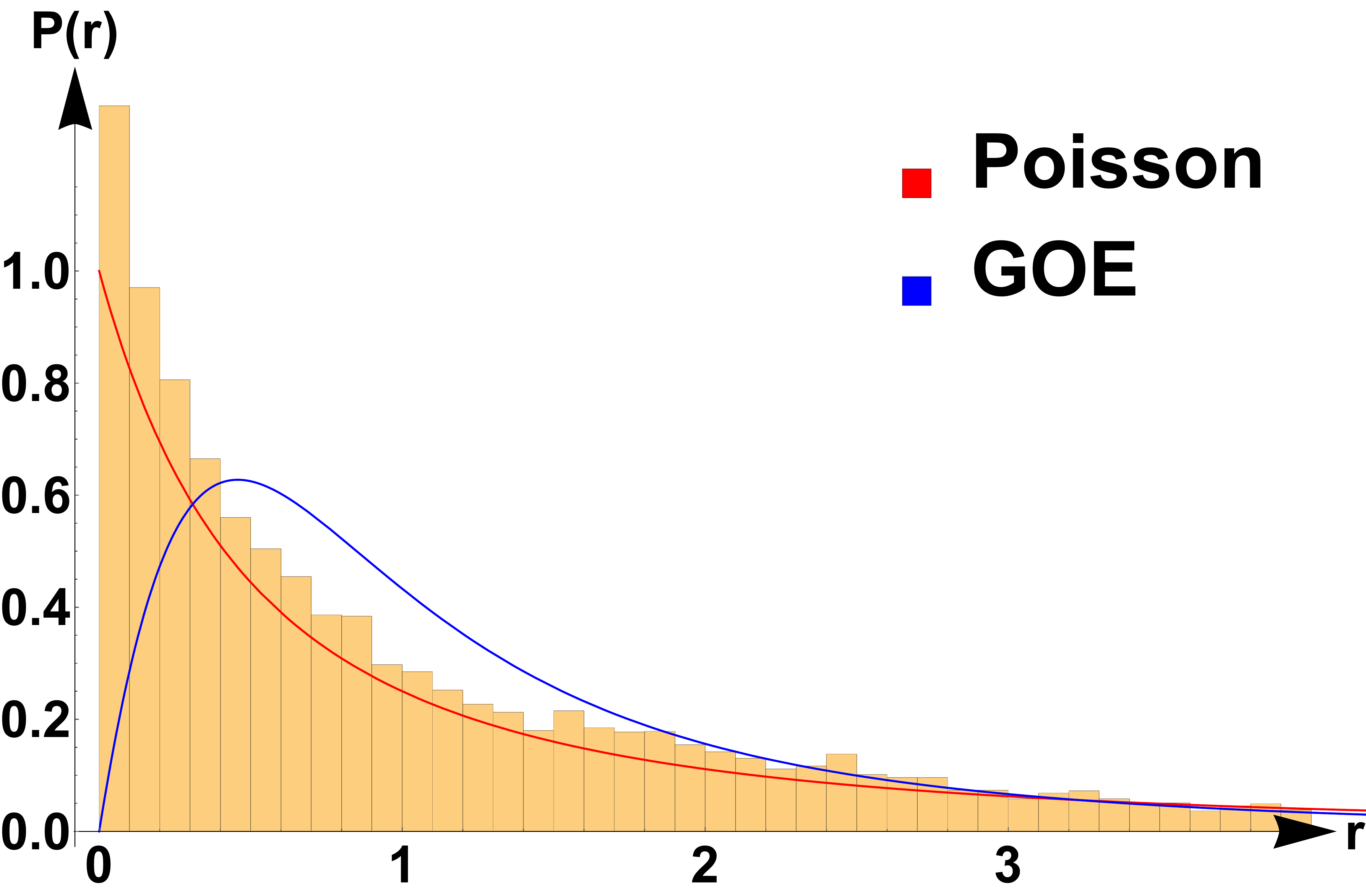}   %
}\hfill
\subfloat[\label{0piext0,k=2,r}]{%
\includegraphics[width=0.3\textwidth]{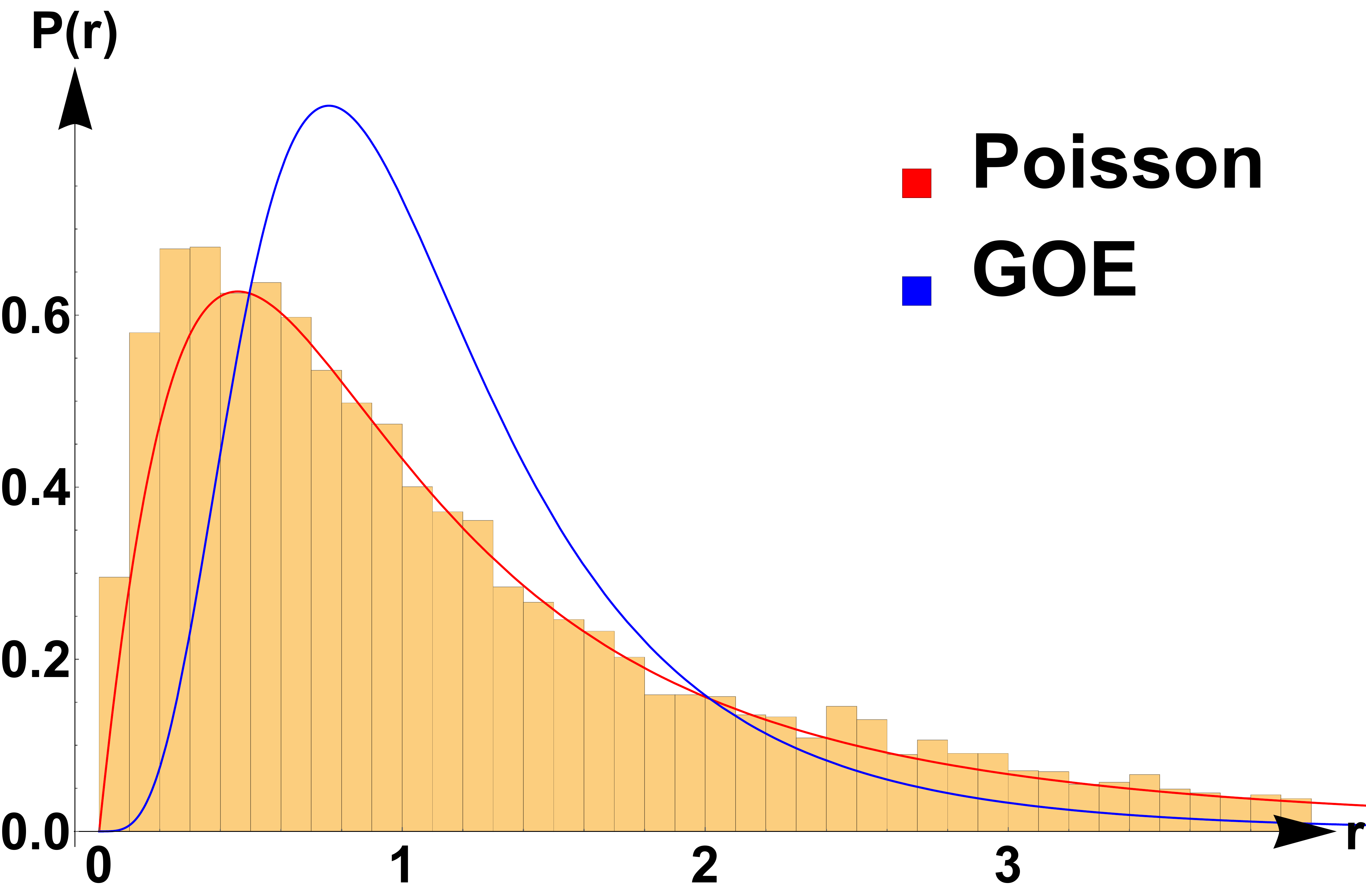}
}
 \\
 \subfloat[\label{0piextpi}]{%
\includegraphics[width=0.3\textwidth]{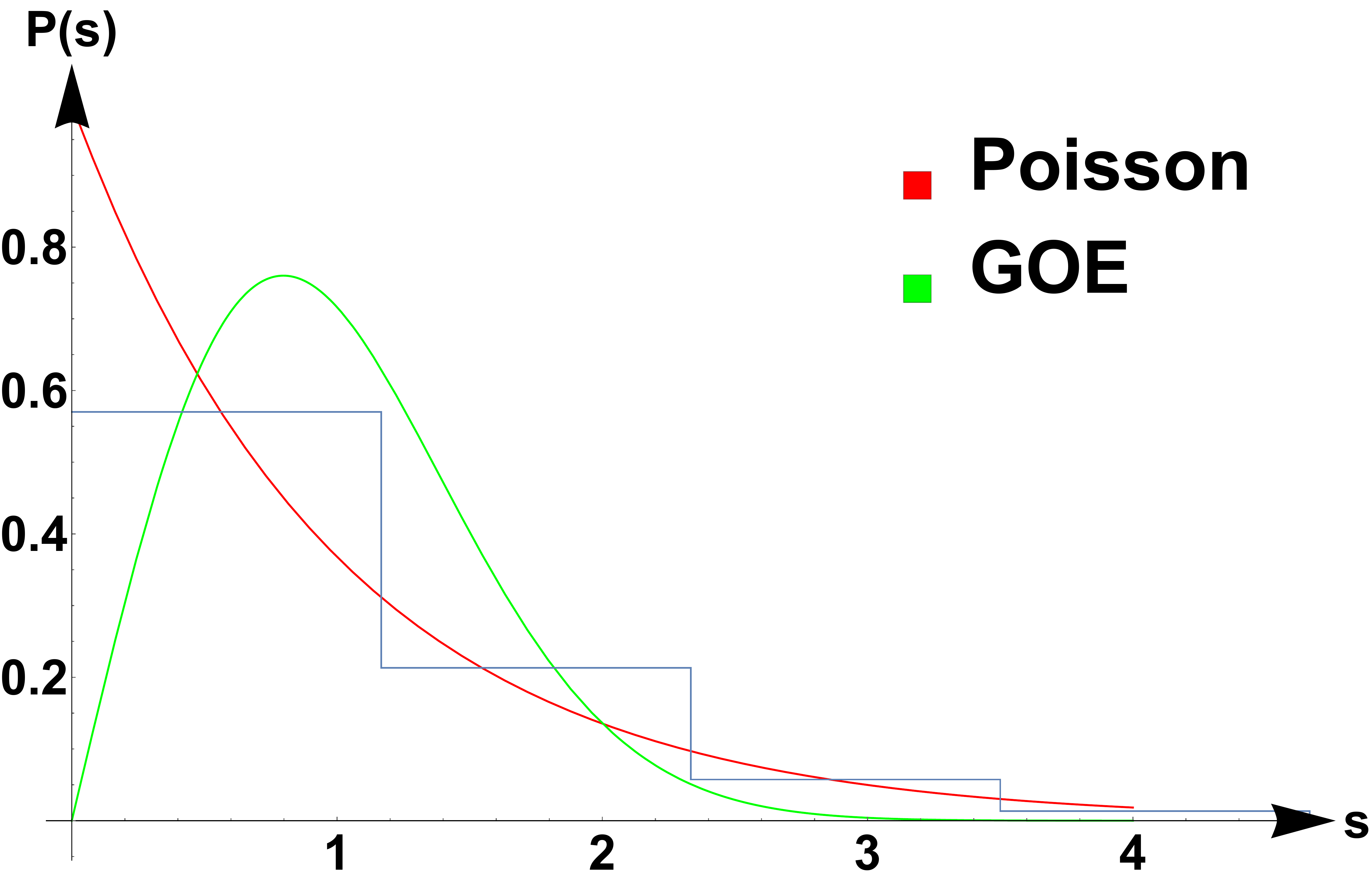}
}\hfill
\subfloat[\label{0piextpi,k=1,r}]{%
\includegraphics[width=0.3\textwidth]{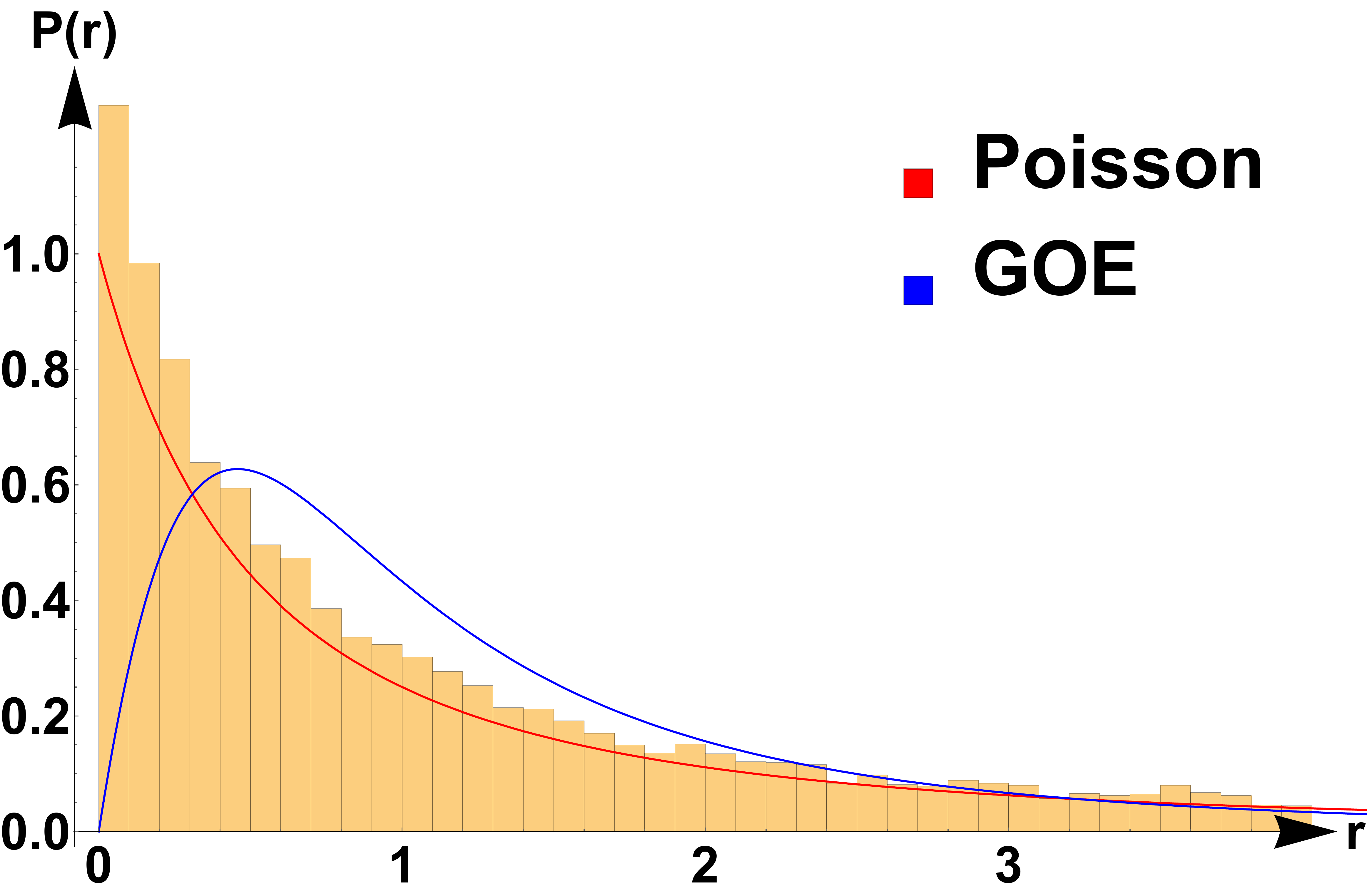}  %
}\hfill
 \subfloat[\label{0piextpi,k=2,r}]{%
\includegraphics[width=0.3\textwidth]{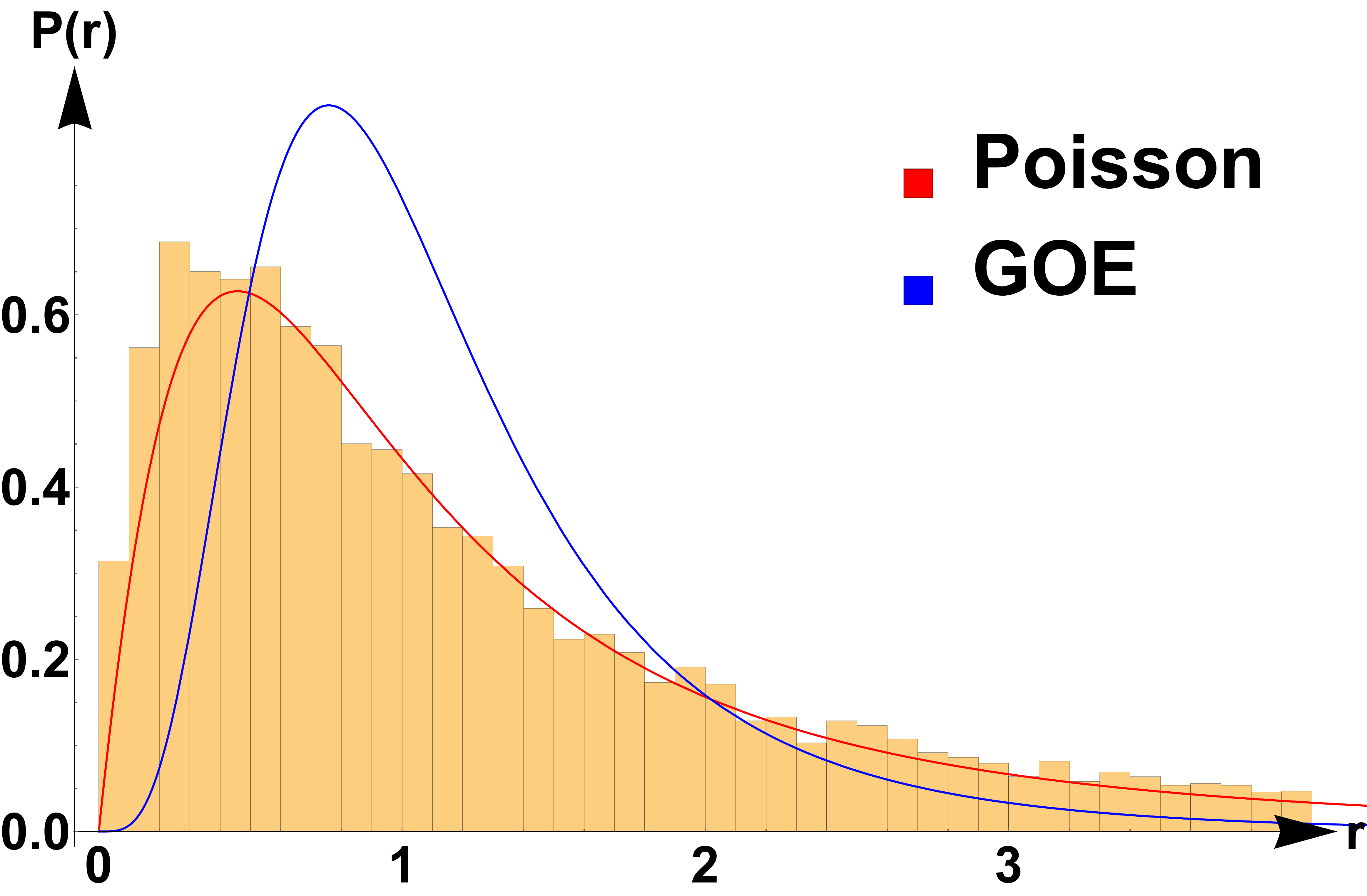}%
}
\caption{In {\ref{0-piclassical0} \& \ref{extpiclassical0-pi}}, we have the phase space plot of Hamiltonian near resonance(1:1) respectively for $\Phi_{ext}=0$ and $\Phi_{ext}=\pi$.  Here,  values of $\Phi_{ext}$ and other variables is taken from \cite{gyenis}. In \ref{1to10energylevel0-piqubitext0} and \ref{1to10energylevel0-piqubitextpi}, first $10$ energy level wavefunctions of the Hamiltonian near resonance are plotted on top of the potential for reference. Wavefunctions are offset by their corresponding eigen-energies. As expected from {\ref{0-piclassical0} \& \ref{extpiclassical0-pi}}, probability is higher in corresponding bounded regions. \ref{0piext0} \& \ref{0piextpi}, the histogram of adjacent spacing is plotted respectively for the $\Phi_{ext}=0$ and $\Phi_{ext}=\pi$. The red line represents the Poisson distribution and green line represents the GOE(Gaussian Orthogonal Ensemble) or Wigner Distribution. In {\ref{0piext0,k=1,r} \& \ref{0piextpi,k=1,r}}, the histogram of ratios of adjacent spacing whereas in {\ref{0piext0,k=2,r} \& \ref{0piextpi,k=2,r}} the histogram of ratios of next nearest neighbour spacing is  plotted respectively for the $\Phi_{ext}=0$ and $\Phi_{ext}=\pi$. The red line represents the Poisson distribution and blue line represents the GOE or Wigner Distribution. Here, we observe that our distributions resembles Poisson distribution for above cases.}
\label{}
\end{figure*}

Quantum circuits can be designed according to the characteristics (location, size etc.) of the desired trapping region with the choice of suitable parameters. The underlying philosophy is to create stability at the edge of chaos (more precisely, non-integrability), ``perfect" (integrable) systems being prone to instability upon any perturbation. The primary resonance condition suggests a relation among the parameters. As illustrated by the examples considered, we have quantitatively shown that a wavepacket will have a negligible tunneling probability to interact with the transmon system situated deep inside the primary island. The quantum circuits can be manipulated nevertheless, or tuned, by introducing time-dependent, well-controlled external probes. For instance, a time-dependent flux can tune a $0-\pi$ qubit in a way that we desire. A systematic gauge-invariant Hamiltonian for such systems coupled to resonators provides an interesting possibility for design of architecture in two and three dimensions. 

To conclude, we have presented a set of criteria for protection of qubits by exploiting nonlinearity, drawing inspiration from well-known instances of stability of classical nonlinear systems.       
\\
R.K.S. would like to thank Nishchal R. Dwivedi and Sandeep Joshi for stimulating discussions and encouragement.

\end{document}